\documentclass{aa}
\include{psfig}
\usepackage{graphics}
\begin{document}
 
\thesaurus{08(09.18.1, 09.09.1 OH 0739-14, OH 231.8+4.2, 03.20.5, 08.09.1 Qx Pup)}
\title{High resolution K$_{S}$ polarization mapping of OH~231.8+4.2 (OH~0739-14) with
SOFI}
\titlerunning{K$_{S}$ polarimetry of OH~231.8+4.2}
\author{N. Ageorges\inst{1,2}
\and
J. R. Walsh\inst{2}
}
\authorrunning{Ageorges \& Walsh}
\offprints{N. Ageorges}
\institute{
Physics Department, 
National University of Ireland - Galway, 
Galway,
Ireland. \\
E-mail: nancy@epona.physics.ucg.ie
\and
European Southern Observatory,
Karl-Schwarzschild Strasse 2, 
D85748 Garching bei M\"{u}nchen,
Germany.
E-mail: jwalsh@eso.org
}

\date{Received / Accepted }
\maketitle

\begin{abstract}
The bipolar reflection nebula OH~231.8+4.2 (OH~0739-14) centred on the
Mira variable star QX Pup has been mapped at K$_{S}$ in linear polarization 
using the ESO SOFI near-infrared multi-mode
instrument. The large scale linear polarization features agree with previous
studies, validating the polarimetric mode of the instrument.
However the high spatial resolution of the polarization map reveals
considerable structure, some of which finds correspondence in the 
surface brightness images. The core is crossed by low polarization filaments;
the northern lobe reveals an axial finger and a rim of very high linear
polarization (upto 80\%) whilst the southern lobe shows less polarization 
contrast. There is a trend in polarization perpendicular
to the bipolar axis in both lobes. The individual linear polarization features 
most probably arise from localised dust structures with distinct orientations 
leading to differing polarization through single scattering. A single scattering
geometric model, consisting of two lobes of ellipsoidal section
and a central dense disk, is considered which can explain the general
features of the intensity and polarization images. 
\end{abstract}

\section{Introduction}
The bipolar nebula OH~231.8+4.2 (IRAS OH~0739-14, sometimes called the 
Calabash nebula) is centred on the Mira variable QX Pup and is a  
source of line and continuum emission at many wavelengths.
It displays strong extended optical and IR structure about 40$''$ in length,
which is very highly polarized
(Scarrott et al \cite{scarr} for the optical; Kobayashi et al. \cite{kob},
Heckert \& Zeilik \cite{heck}, Shure et al. \cite{shur}, Kastner \& Weintraub
\cite{kawe} for the IR). At the extremity of the 
southern bipolar lobe, there are high velocity emission line knots,
similar to Herbig-Haro objects (Reipurth \cite{reip}). 
The central variable star has an amplitude of $\sim$2 mag., a period of 
$\sim$700 days (Kastner et al. \cite{kast92})
and a spectral type of M9~III (Cohen \cite{coh81}; Cohen et al. \cite{coh85}).
The star is not directly visible in the optical on account of a waist
of high extinction, but dominates the structure at longer wavelengths 
(Woodward et al. \cite{wood}). Kastner et al. (\cite{kast92}) have resolved the star 
from the surrounding nebula at K and L and deduce that it is a typical Mira
variable. The object was first discovered on account of its OH maser 
emission (Turner \cite{turn}) and mm wave CO mapping has shown a high velocity
outflow oriented at $\sim$40$^\circ$ to the plane of the sky with deprojected
velocities upto 330 kms$^{-1}$ for the southern lobe and 90 kms$^{-1}$ for
the northern lobe (Alcolea et al. \cite{alco}). It is generally considered
to be a transition object between the Asymptotic Giant Branch (AGB) and the 
Planetary Nebula phases in which the high velocity stellar wind is blowing out
the AGB dust envelope. The estimated mass of the molecular and dust envelope
is estimated to be $\sim$0.5 - 1M$_{\odot}$, consistent with a main sequence
mass $\sim$3 M$_{\odot}$ (Kastner et al. \cite{kast98}).

  Aside from its great astrophysical interest as a short-lived
proto-planetary nebula, OH~231.8+4.2 is a useful test source for 
polarization measurements. From the optical to the L band it possesses
high linear polarization (upto 50\% in large apertures, Shure et al. 
\cite{shur}) and so is a useful polarization calibrator. It cannot be 
stated with any certainty that it possesses the most important attribute 
for a calibrator - that the polarization is non-variable, but the high 
polarization makes it useful as a calibrator for the polarimetric 
efficiency and for position angle determination. In addition, since  
the polarization pattern is centro-symmetric along the bright
lobes (although not in the equatorial direction) and extended over a 
length of about 50$''$, it provides an ideal source to confirm any 
instrumental depolarization effects. In consequence OH~231.8+4.2 was 
selected as a test source for the commissioning of the SOFI 
polarimetric mode. The quality of the instrument and the observations 
proved to be excellent and yielded the highest resolution near-IR 
polarization map of this source.

   The instrumental set-up and the observations are detailed in 
Sect. 2 and the polarization results presented in Sect. 3. 
The spatial polarization structure is discussed in Sect. 4 together
with conclusions about the geometry of the elongated dust cloud and
the scattering properties of the grains.

\section{Observations}

\subsection{SOFI}
SOFI (Son OF Isaac, Moorwood et al. \cite{moor}) is an infrared spectrograph and 
imaging camera installed at the ESO 3.5m NTT. It has a focal reducing 
camera and is capable of performing near-infrared imaging, 
spectroscopy with grisms and polarimetry. Imaging can be achieved with 
3 different plate scales (0.144, 0.276 and 0.292$''$ per pixel) using 
broad and narrow band filters in the wavelength range 0.9 to 2.5$\mu$m.
The detector is an Hawaii HgCdTe 1024 $\times$1024 pixels array, with a quantum 
efficiency of 65\% (Moorwood et al. \cite{moor}). 
Imaging polarimetry is possible over the near-IR range using a Wollaston prism 
installed in the grism wheel and a special mask in the aperture wheel. The
mask has three focal plane apertures with separation chosen so that the 
image pairs resulting from the Wollaston do not overlap on the detector.
Broad and narrow filters may be selected together with the Wollaston. 
Using the Wollaston and aperture mask three pairs of images are collected;
one pair for the object and two pairs for offset sky. 
The Wollaston outputs two divergent beams with the E-vector oriented
parallel and perpendicular to its axis; 
the difference between the two images allows  
one Stokes parameter to be determined. The size of each strip is about 
1023 by 145 pixels, separated by about 8 pixels.

\subsection{Observational set-up}
The observations of OH~231.8+4.2 were performed on 1997 August 8 
using the 0.292$''$ pixel scale. Data have been acquired with a narrow 
band filter of 2.162$\mu$m central wavelength and 0.275$\mu$m width
(K$_{S}$) using double correlated sampling with 60 samples. 
Double correlated reading measures the signal as the difference between 
the reset voltage of a diode and the post exposure voltage; the noise 
from thermal capacitance changes are eliminated as a result. 
For each of two position angles of the Wollaston, separated by 45$^\circ$ 
measurements were repeated three times with an exposure time of 70.9s.  
Between each acquisition the position of the nebula on the chip was displaced, 
to avoid using exactly the same detector position. This also has the advantage 
that no extra sky data needed to be acquired. By subtracting one data set from the 
other, the sky contribution can be removed. The sets of three images
allowed one of the Stokes parameters, U or Q, to be measured; the set
taken at a position angle rotated by 45$^\circ$ allowed the other Stokes 
parameter to be determined.

\section{Reductions and results}

\subsection{Data reduction}
\label{reduction}

The six regions on the detector, one for the object (O) and two for the 
sky (S$_1$, S$_2$), each of them with two transmissions of the Wollaston
(referred to as O (ordinary ray) and E (extraordinary ray)), 
at which data have been acquired, are called hereafter On$^o$,On$^e$, 
Sn$_1^o$,Sn$_1^e$ and Sn$_2^o$,Sn$_2^e$, where n refers to the
repeat observations taken with the object shifted on the 
detector. There are two ways in which this data can be
treated: either the mean sky can be formed from the offset sky
images from (Sn$_1^o$, Sn$_2^o$) and (Sn$_1^e$, Sn$_2^e$) 
and subtracted from the corresponding On$^o$ and On$^e$;
or the repeat (`nodded') observations on the object images 
can be subtracted (i.e. O1$^o$ and O2$^o$), for the same Wollaston 
transmission. The latter procedure has the advantage that the
sky is subtracted from the same detector pixels as the object and
the procedure should be insensitive to pixel-to-pixel variations;
however it has the potential disadvantage that for extended objects 
there may be stars in the same position as the object in shifted
frames. Comparison with the result using the mean sky formed
from the offset images can be used to ensure that no stars are in fact 
subtracted from the nebular image. The nodding procedure was followed 
to form the nebular images
before combination, with the modification that the image to be subtracted
was formed from the mean of the other two (i.e. O1$^o$ minus
mean of O2$^o$ and O3$^o$). Thus a total of 12 object images
were processed (three positions of the object on the array, each with O 
and E and with the instrument rotated by 45$^{\circ}$). 
Flat field exposures were taken using dome illumination with the same
filter and Wollaston prism. Flat field images were created after extraction 
of similarly defined areas (same size and position) from the original 
dome flat fields and normalization. It was found that the flat 
fields implied small ($^<_\sim$5\%) but significant changes in the 
pixel-to-pixel response.  
 
The basic data reduction steps were performed with the `eclipse'
package (Devillard \cite{nico}).
The source frames were first sky subtracted, flat fielded and then bad pixel
corrected. Sky data, determined as explained above, has been used to 
derive bad pixels maps, via a  `median threshold' method which detects the 
presence of spikes above or below the 
local mean in the individual image. Assuming the signal is smooth enough, 
bad pixels are found by computing the difference between the image and its 
median filtered version, and thresholding it. The algorithm used to clean 
out a frame from bad  pixels replaces bad pixels  
by an average of the valid pixels from of the 8 closest neighbours. 

Combining the object images from one set of O and E ray Wollaston data
allows one Stokes parameter to be derived. 
One of the data sets must be rotated by 45$^\circ$ in order to
determine the linear polarization from two Stokes parameters.
To avoid any artefact that the rotation algorithm might create in the structure
of the data, it was decided to rotate the images acquired with the
Wollaston at 0$^{\circ}$ by +22.5$^{\circ}$ and the ones with a position
angle rotated by 45$^\circ$ by
-22.5$^{\circ}$ respectively, so that both data sets are equally affected
by this operation. The final results were compared with those produced by
rotating one set by 45$^\circ$; no significant differences were
noted but the more conservative procedure was followed.

To estimate the spatial variation of the polarization, the Q and U Stokes
parameter images had to be shifted to the same position. For this
purpose, the centroid position of the star at position $\Delta$X =
+16$''$ and $\Delta$Y = +9$''$ from the core of OH~231.8+4.2 
(Fig. 1) was calculated to
within a tenth of a pixel, and used to determine the necessary image
shift. The IRAF `imshift' procedure was used,
ensuring that all images are at the same position within two tenths of a
pixel. Since, however, combining images with a rotation of 45$^\circ$ 
is the worst case situation as far as sampling is concerned, the images were
first pixel replicated to 4$\times$4 smaller pixels. Then the images
were aligned using this star and rotated to a common position angle. 
The final images were then 4$\times$4 binned back to the original pixel 
scale. A seeing of 1.0$''$ (3.5 pixels) was measured during the observations,
although the data taken at a PA of 45$^\circ$ had slightly worse seeing.
After the rotation, alignment and summing of the three data sets acquired with 
one given polarization (O or E ray for the Wollaston at 0 or 45$^{\circ}$), 
the degradation of FWHM was found to be minimal (less than 0.1$''$). 

Polarization maps have been created both for flat-fielded and non
flat-fielded data using the imaging polarimetry package IRAF.stecf.impol. 
The SOFI detector is known to suffer from 
a signal dependent bias so that the absolute level of the flat field
exposure can induce artefacts into the flat field. The flat fields
were fairly flat in signal over the exposed regions of the chip 
(rms $\leq$5\%), so the induced effects should not be large. The
primary aim of the flat fielding was to reduce pixel-to-pixel
scatter. Examination of the final images showed that the
flat-fielded images were indeed slightly smoother than the non flat-fielded
ones. Comparing the flat-fielded and non flat-fielded polarization maps
showed that there were no features, above the noise, detected in one 
image which were not present in the other.

 The instrumental polarization had not yet been determined prior to
the observations of OH~231.8+4.2. A series of observations of the
unpolarized standard star HD~94851 (B polarization 0.057\%, Turnshek
et al. \cite{turns}) were made rotating the instrument
by 22.5 degrees from 0 to 180$^\circ$ in steps of 22.5$^\circ$.
All the O and E ray data at all position angles were reduced together
for both flat fielded and non-flat fielded data frames. The results
were obtained fitting a cosine 2$\times$$\theta$ curve to all the data 
points using least squares. For the flat fielded data the instrumental
polarization at K$_{S}$ was found to be 1.03$\pm$0.07\% at 142.2$\pm$1.8$^\circ$.
This determination assumes that the polarization of HD~94851
at 2.1$\mu$m is zero; the value is not known but is probably less than
0.01\% assuming a normal interstellar polarization curve. This value
of the instrumental polarization was used to correct the 
linear polarization and position angle of the OH~231.8+4.2 data.

\subsection{Results}
\label{results}
The total intensity K$_{S}$ map formed by summing all the images in the two Wollaston 
sub-images and two orientations, aligned, as described above, on the star 
to the south-west, is shown in Fig. 1a in a logarithmic plot. Part of the 
extended nebulosity to the S, containing Herbig-Haro objects (Reipurth 
\cite{reip}) was out of the imaged area. The sky subtracted and flat fielded 
images in the two Wollaston sub-images and at the two rotator angles were
combined to form images and errors in Q and U Stokes parameters using
the IRAF.stecf.impol package. The errors were propagated from the original 
images assuming a conversion of 5.9 e$^-$ per ADU and a read-out noise of 
12 e$^-$ per pixel. The linear polarization, the corresponding error map, 
the position angle and error image 
were determined. The polarization was corrected for non-negativity bias (e.g.
Wardle and Kronberg \cite{wakr}) and the position angle errors determined from 
the distribution  given by Naghizadeh-Khouei \& Clarke (\cite{nag}). No
correction was applied for intervening interstellar polarization. Fig. 1b shows the 
image of the linear polarization in OH~231.8+4.2 in an unbinned form
(cut-off in polarization error is 10\%). Fig. 1c 
shows the 2$\times$2 pixel binned linear polarization map. The cut-off in 
polarization error is 10\% in this map. 
A wealth of structure in the polarization map is evident
attesting to the fact that it is not a simple centro-symmetric
reflection nebula. The polarization structure is discussed in Sect. 4.

\begin{table*}
\caption[]{Aperture polarimetry of OH~231.8+4.2}
\begin{flushleft}
\begin{tabular}{lccrrrr}
Aperture & Offset$^\dagger$ & Aperture    & Linear~~~     & Poln.~~~~~ & \multicolumn{2}{c}{Literature$^\ast$} \\
Name     & ($''$)~~         & Diam ($''$) & Poln. (\%)~ & PA ($^\circ$)~~~ & Poln (\%)  & PA ($^\circ$) \\
\hline
C      &  0.0,0.0       & 4.0 & 23.44$\pm$0.03 & 108.69$\pm$0.03 & 24.4$\pm$0.5 & 110 \\
A      & $+$2.5,$+$7.0  & 4.0 & 50.84$\pm$0.08 & 108.32$\pm$0.05 & 50.0$\pm$0.7 & 112 \\
E      & $-$2.5,$-$7.0  & 4.0 & 46.15$\pm$0.06 & 124.54$\pm$0.04 & 49.8$\pm$2.4 & 117 \\
\#1    & $+$5.0,$+$13.8 & 3.5 & 39.59$\pm$0.10 & 108.40$\pm$0.07 &  & \\
\#2    & $+$0.7,$+$1.7  & 2.3 & 38.90$\pm$0.04 & 104.95$\pm$0.03 &  & \\
\#3    & $-$0.8,$-$2.1  & 4.1 & 25.48$\pm$0.02 & 109.74$\pm$0.02 &  & \\
\#4    & $-$0.3,$-$2.1  & 1.8 & 29.33$\pm$0.04 & 101.31$\pm$0.04 &  & \\
\#5    & $-$4.1,$-$10.2  & 2.3 & 48.57$\pm$0.21 & 126.08$\pm$0.12 &  & \\
\#6    & $-$7.5,$-$18.5  & 2.9 & 50.51$\pm$0.74 & 116.93$\pm$0.42 &  & \\
Star X &                 & 2.9 &  2.65$\pm$2.63 & 179.84$\pm$31.9 &  & \\
\hline
\end{tabular}
\end{flushleft}
$^\dagger$ Offsets in RA and DEC, relative to the the position of the central star:
East $+$, South $-$ \\
$^\ast$ Literature values from Shure et al (1995) in K band. \\ 
\end{table*}

  In order to compare the SOFI polarimetry with previous work, the signal in a
number of apertures was summed and the polarization and position angle
values determined. Measurements were made in three circular apertures 
at the centre, at 2.5$''$E,7.0$''$N and at 2.5$''$W,7.0$''$S (respectively
the apertures C, A and E of Shure et al. \cite{shur}) and are listed in Table
1. Fig. 2 shows a sketch of the appearance of the nebula in K$_{S}$ 
with the positions of these three apertures shown by dotted circles.
In addition the linear polarization at six other positions ofinterest, 
whose exact positions are detailed in Table 1, was determined for the northern
crescent knot, the fainter central knot, the whole of the brighter S knot,
the peak of the southern knot, the knot to the south-west and the extreme
southern knot. Aperture diameters, selected to include these features, are 
listed in Table 1 and the aperture positions are also shown in Fig. 2.
The linear polarization and position angle values at K from 
Shure et al. (\cite{shur}) are also listed in the last column of Table 1. The 
agreement is generally excellent: for position C the linear polarization
and position angle agree within errors; for position A (the brightest
region) the agreement in polarization is fair; for position E, 
the error bars are larger since this is the faintest region of the three, 
and the position angle is discrepant. The small discrepancies may be partly
attributable to differences in the exact positioning of the different 
apertures. The last item in Table 1 is the alignment star to the SE (upper
right in the figures); it has a polarization consistent with zero to within
the errors. 

Given this good agreement between the aperture polarimetry results presented
in Table 1 and those of other observers, the results of the imaging 
polarimetry with SOFI are considered to be reliable, although a full calibration
of the instrument has not yet been performed. It is concluded that 
there is no large correction for polarizing efficiency or position angle 
to be applied. Although this source is certainly useful as
a high polarization calibrator, the position and size of the 
aperture must be carefully chosen otherwise discrepant results
can be obtained. 

\begin{figure*}
\centerline
{\vbox{\resizebox{0.67\hsize}{!}{\includegraphics{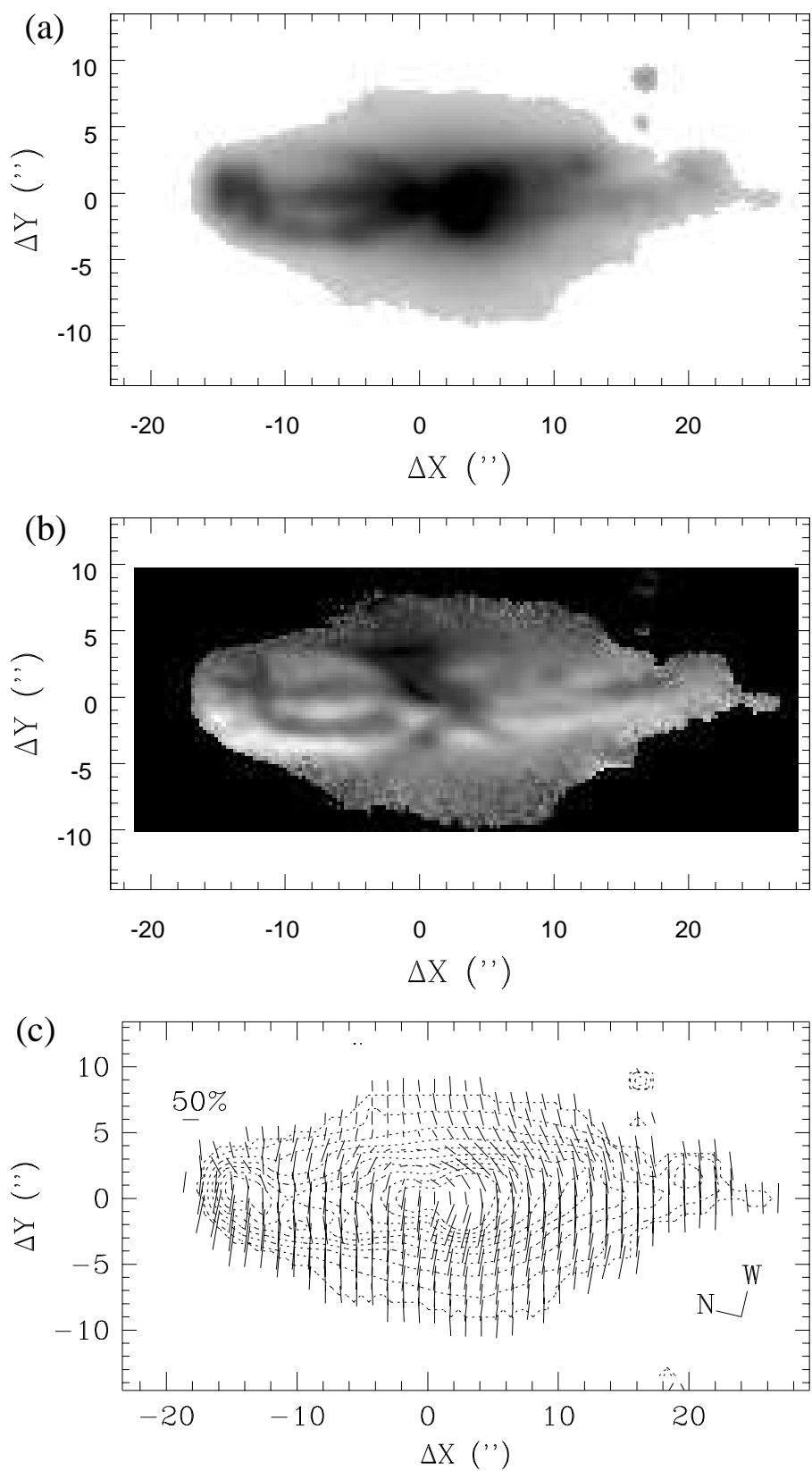}}}}
\caption{(a) K$_{S}$ total (relative) intensity map of OH~231.8+4.2 is displayed 
in a logarithmic plot. The data are at the original pixel 
size (0.292$''$);
(b) greyscale plot of the degree of linear polarization map of OH~231.8+4.2 is 
shown, with a range from 0 (black) to 90\% (white);  
(c) polarization vector map of OH~231.8+4.2 derived from the K$_{S}$ images is shown. 
Logarithmic contours of the intensity map are overplotted (dashed lines). 
The vectors have been binned 2$\times$2 pixels before plotting
and only every 2nd vector is plotted for clarity.
}
\label{pols}
\end{figure*}

Fig. 3 shows the profiles of total intensity, linear polarization and 
position angle along the long axis of the nebula (PA 20$^\circ$); each point
corresponds to three pixels binned.  
Offset 0 corresponds to the position of the central star as estimated by
comparison with the images of Kastner et al. (\cite{kast98}). This central position 
also displays the lowest value of linear polarization and the position angle
shows an abrupt change across this region. Error bars are shown for the 
polarization and position angle in Fig. 3.

\begin{figure}[htb]
\vspace{1.4cm}
\centering
\resizebox{0.72\hsize}{!}{\includegraphics{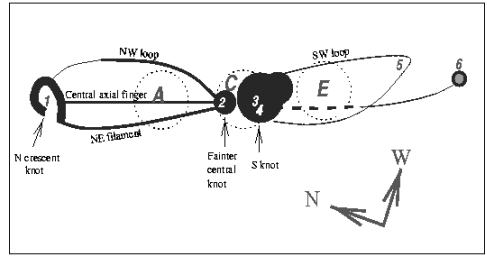}}
%
\caption{A sketch of the appearance of OH~231.8+4.2 at K$_{S}$ is shown
with the features referred to in the text labelled. Apertures A, C and E
(Shure et al. \cite{shur}) are indicated by dashed circles and the centres
of six other apertures at which the linear polarization was determined
(Table 1) are numbered.}
\label{sketch}
\end{figure}

\begin{figure}
\centering
\resizebox{\hsize}{!}{\includegraphics{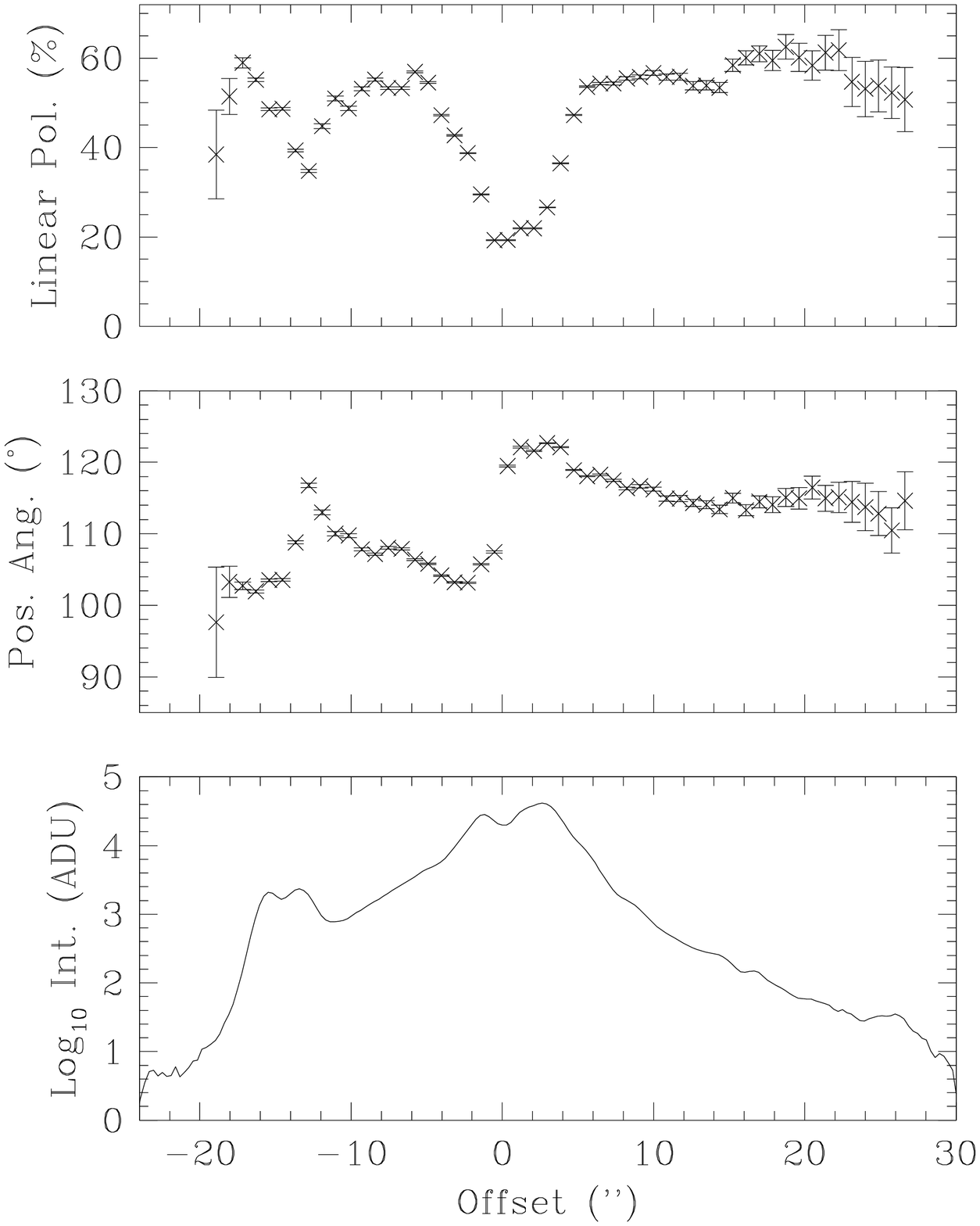}}
\caption{3$\times$3 pixels binned cuts through the long axis of the 
nebula at absolute PA of 20$^\circ$ (i.e. the horizontal in Figs. 1a-c)
are illustrated. The top plot shows the linear polarization, the middle
plot the polarization position angle and the bottom plot 
the log of the (relative) total signal (Stokes I). Offset 0 corresponds
to the position of the central Mira star.}
\label{cuts}
\end{figure}

\section{Discussion}

\subsection{Intensity and polarization morphology}

The K$_{S}$ total intensity morphology shown in Fig. 1a consists of 
two lobes which however are substantially different in appearance. 
The sketch of the nebula in Fig. 2 indicates the morphological features
distinguished here. The northern lobe, which is tilted towards the 
observer (from the plane of the sky) by about 40$^\circ$ (e.g. Cohen 
et al. \cite{coh85}; Kastner et al. \cite{kast92};
Scarrott et al. \cite{scarr}) has a central finger and
two curling lobes emerging from the core which hides the 
Mira star QX Pup behind a dense disc of dust (Kastner et al. \cite{kast98}).
The finger and north-eastern filament meet in the northern crescent shaped 
knot (\#1 in Table 1).  The southern lobe is less sharply defined; 
the eastern loop is not present and the western one is
composed of several knots. The high resolution achieved without 
adaptive optics can be appreciated by comparison with the
K band AO imaging of Kastner et al. (\cite{kast98}). The features on the AO 
image are not much sharper than in Fig. 1a and the lobes
are not strikingly composed of many resolved knots.

Kastner et al. (\cite{kast92}) have also published a K band polarization 
map (their Fig. 2). There is broad agreement with their Fig. 2 
but the SOFI map has higher spatial resolution and signal-to-noise. The 
linear polarization images in Figs. 1b and c show important features in 
common with the total intensity map which bring a wealth of new information 
on the structure of the nebula and the nature of the dust features.
The new polarization features can be summarized as follows:\\
the central finger in the N lobe has a higher polarization than the 
immediate surroundings, by about 20\% (Figs. 1b and c); \\
the northern knot has low polarization (offset -13$''$ in Fig. 1b and c)
and the position angle is distinct across the knot; \\
there is a line of lower polarization running diagonally across the
central region (Figs. 1b and c); \\
there is an outer halo of lower polarization (though still $\sim$30\%) 
around the nebula (Figs. 1b and 1c); \\
the polarization increases from west to east across both lobes, but most 
noticeably for the northern lobe (Fig. 1b,c). \\
Comparing Figs. 1a and 1b it is clear that the finger and loops visible 
in total intensity in the N lobe are also more strongly polarized, 
whilst in the southern lobe (tilted by $\sim$40$^\circ$ away from the 
plane of the sky) there is no such clear 1:1 correspondence. 

The most remarkable features are the elongated very high polarization 
features in the north lobe. Along the central finger the polarization
increases from 48 to 71\% (at 5.3$''$ from the central star) before
declining again to values around the average for the N lobe
(40\%). The eastern filament in the N lobe has its peak value of
80\% at an offset of 10.5$''$ from the centre; the position angle
of this entire feature is 20-30$^\circ$ greater than the surroundings. 
The other dominant
feature in this lobe, the knot at -15$''$ offset, however has low
(30\%) polarization and the position angle is distinct and close
to the value for the S lobe. This difference may be due to intrinsic emission
diluting the polarization or, more probably, an orientation closer to
that of the S lobe (i.e. in the background). The south lobe has 
a feature along its length with linear polarization decreasing from
70\% to 60\% although the collimation of this feature is much less clear
than for the N lobe and there is no distinction in position angle relative
to the surrounding lobe. This high polarization feature in the S lobe 
is towards the eastern side and is more similar to the high polarization 
eastern edge to the N lobe, but does not correspond clearly 
to a filamentary feature on the intensity image. The lower contrast of
features in the S lobe may be caused by obscuration by the equatorial disc. 
The knot at +11$''$ offset on the western side of the S lobe with 35-40\% 
polarization parallels the low polarization crescent knot in the N lobe but is 
less striking in its brightness and polarization minimum. 

The almost circular cloud of lower polarization values surrounding
the bipolar nebula are interpreted as belonging to the disc. The
low polarization on the front side of this disc is clearly seen 
as the minimum a few arsec to the south of the position of the central star
in Figs. 3 and 1b and c. Note also how the position angle of polarization
changes across this region (Fig. 3 offsets 0 to +8$''$).
This coincides with the brightest spot in Fig. 1a. If this 
was single scattering from the surface of the disc, the polarization 
would be expected to be higher than for the northern lobe at a similar 
offset. Since this region is offset from the position of the 
central star, the scattered flux cannot be diluted by
direct radiation, so that multiple scattering must be occurring.
The minima in polarization around the central star
can be attributed either to viewing directly along the flaring edge 
of the bipolar lobe as it emerges from the thick disk, or to 
multiple scattering at the dense interface region between the equatorial 
disc and one of the bipolar cavities. The gradual increase in polarization
along the bipolar axis into the south lobe (Fig. 3) can be 
explained by an increasing mix of scattered flux from the bipolar lobe
over that of the disc as the line of sight optical depth through the
disc decreases outward. 
 
The lower polarization region around the nebula that shows the bulk of the
polarization vectors are aligned radially, rather than normal to the direction
to the central star, as first noted by Dougados et al (\cite{doug}).
This has been traditionally ascribed to multiple scattering in an optically
thick dusty disc (e.g. Kastner \& Weintraub \cite{kawe}; see also
Bastien \& M\'{e}nard \cite{bast}). The possibility 
that there is substantial flux scattered off the lobes reflected from the 
surface of the disc seems less likely; the scattering geometry would 
produce a range of position angles with only two scatterings and so is probably 
not a significant effect. However the constancy of the position angle over 
large areas of the disc, in particular to the east (Fig. 3), suggest a large
scale organization of grain alignment. The degree of multiple scattering 
is primarily dependent on the optical depth, which should vary with position 
around the disc, so may not be able to produce such a coherent pattern of
polarization vectors. Bulk alignment of (non-spherical) grains by a 
frozen-in magnetic field might however be capable of explaining the observed
configuration (e.g. Hildebrand \cite{hild}).

\subsection{Dust features}
The high polarization, together with the stronger K$_{S}$ signal, over the finger 
and the eastern loops in the northern and southern lobes requires to be 
explained. The higher
K$_{S}$ flux must be attributable to high column depth of dust since it cannot 
be explained together with increased polarization: polarization increases
while scattered flux decreases with increasing scattering angle for 
interstellar dust scattering situations (e.g. Mathis \cite{math}).  Two 
possibilities are considered for the high polarization features: \\
\begin{itemize}
\item the high polarization is caused by differing scattering angles. The higher
values correspond to values closer to 90$^\circ$ scattering where polarization,
at least for Rayleigh scattering, is maximum. Since the N lobe is tilted towards
the observer it would imply that the finger and the eastern wall
are tilted more away from the observer. To avoid the over-complex view that the 
lobe is composed of several components, all with different tilts, 
the lower polarization regions could represent an integration along the line of
sight of a range of scattering angles. The lower polarization of the western edge
then indicates a smaller scattering angle and hence an east-west tilt of a lobe 
of elliptical section. The central finger would have to be at a rather different tilt 
from the lobe axis for its polarization to be dissimilar to the 
surroundings; \\
\item the dust in these features is indeed different from that in the bulk of the lobe.
In order for the polarization to be higher then the particles would have to be 
smaller. If the lobes are hollow then it is difficult to see why one edge 
of the lobe should happen to have differing particles to those averaged 
through the lobe. The hypothesis of differing particles can only realistically 
be applied to the finger.
\end{itemize}

  The first suggestion could explain the differing polarization of the
loops, as compared to the material within the lobe, by having the lobe
more elongated in the line of sight. The tilt in polarization
perpendicular to the projected major axis of the nebula could be explained
by an elliptical section with the eastern side more twisted away from 
the observer. Fig. 4 shows a sketch of this configuration. A 3-D dust
nebula with this shape was generated and scattered intensity and polarization 
maps were computed (cf. the polarization models for $\eta$ Carinae in Meaburn et al
\cite{meab}). Rayleigh scattering was employed, although problems with
the presence of small grains have been suggested by Shure et al. (\cite{shur})
on account of the similar optical depth with wavelength in the near-IR. 
It was found from the scattering model that the twisted ellipsoidal 
geometry produces a reverse (i.e. increasing east-west)  trend in the polarization
across the south lobe. Thus there is a suggestion that the southern lobe
is twisted in the opposite sense about the long axis to the northern lobe.
However given the irregularity of the southern lobe this is no more
than a suggestion. A polarization map formed from the 3-D model is shown in 
Fig. 5 where the nebula was tilted at 40$^\circ$ to the plane of the sky
and the lobes of elliptical section were twisted about their major axis by 
40$^\circ$. A qualitative match to the fall-off of intensity along the axis 
(c.f. Fig. 3) is achieved for a density gradient of r$^{-0.5}$ along the
lobes. A small central disc, whose thickness is however 40\% of the length of each
lobe, and has a density gradient $\propto$ r$^{-2}$ has been included. The 
model manages to reproduce a number of the large scale features such as the
bright loops bounding the lobes, as well as several salient features such as the 
low polarization filament to the west of the centre and the east-west slope
in polarization across the N lobe. The slope in polarization of the S lobe is
in the reverse direction to that observed and the lower polarization and
radially oriented position angles of the material in the equatorial `disc'
are not matched by this model. Further modelling with multiple scattering for
the central disc was not able to produce the observed pattern of radially
oriented polarization vectors.

  None of these alterations in geometry can simply explain the central finger and
it seems most probable that it is indeed a distinct feature with a different 
tilt or composed of smaller grains. However the trend in polarization along the 
axis of this feature could more easily be attributed to a curve of the feature 
from close to the plane of the sky (i.e. towards the backside of the lobe) 
towards the line of sight, 
rather than a continuous change in grain properties. Given the
fast outflow from the star (the CO data show a linear increase of radial
velocity with deprojected distance of 8.9kms$^{-1}$ arcsec$^{-1}$, Alcolea
et al. \cite{alco99}) and its variability, an outflow with temporally
varying dust properties  
could perhaps not be excluded. Polarimetry at different wavelengths for
this feature could test if the grain properties differ from the surroundings.
Clearly high spatial resolution spectra
are required to measure radial velocities along this feature in order to
distinguish a differing outflow direction. However this is not simple
if the feature is only visible in scattered light since polarization
velocity modelling is required (see Meaburn et al. \cite{meab}). Emission
lines such as [N~II] and [S~II] from shocked gas would be a discriminating probe
of the geometry of this feature. At present there is no evidence that this is
a true jet until high spatial resolution imaging or spectrometry reveal intrinsic
forbidden line emission.

The N lobe finger and the eastern loops in both lobes have such a
remarkably high polarization which implies very stringent limitations to
these features. They must be highly collimated otherwise the change in
scattering angle across them would dilute the polarization; the 
dust must consist of very small particles so that only Rayleigh
scattering is occurring with no admixture of large grains;
there must be very little local emission from the grains or gas.
The method to produce bipolar outflows is usually taken as
interacting winds (e.g. Icke et al. \cite{ipb}) or magnetic fields
(Garcia-Segura et al. \cite{garc}). The presence of highly collimated
dust, which is considered to have been ejected by the slow, dense wind
in the Asymptotic Giant Branch (AGB) phase has not previously been explicitly
considered in the hydrodynamic models. Since the central star of
OH~231.8+4.2 is still too cool to ionize the nebula, the shaping of the
dust features must have occurred by compression and/or erosion by the
cool fast wind, which is deduced from the high molecular line outflow
velocity (Alcolea et al \cite{alco99}). Alternatively the red giant wind 
is itself asymmetric and the
collimated features arise in wind collimation at the star surface.
Garcia-Segura et al. (\cite{garc}) have shown that stellar
magnetic fields of hundreds of Gauss together with stellar
rotation rates close to critical rotation speed can produce highly
collimated outflows from AGB and post-AGB stars. The very high polarization
values, implying strong confinement, strongly suggest that magnetic
fields or instabilities could play a part. The HST press release image of
OH~231.8+4.2 (PRC 99-39) shows a series of undulations along the N finger and the 
NE filament which present evidence of clumping that could arise from collimation
of magnetic origin. If magnetic collimation does occur, then there may be 
microscopic alignment of grains, which could contribute to the very high
linear polarization values observed, and would also produce measurable
circular polarization. Such microscopic alignment of grains is also suggested
for the equatorial disc on the basis of the large scale pattern of
radially aligned polarization vectors. Mid-IR polarimetry would allow
direct detection of polarized emission from aligned grains, as found
for the Homunculus nebula around $\eta$ Carinae (Aitken et al. \cite{ait95}) 
for example.

\begin{figure}[htb]
\centering
\resizebox{0.8\hsize}{!}{\includegraphics{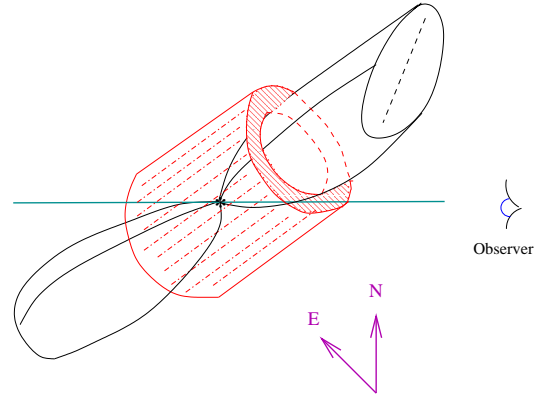}}
\caption{3D representation of OH~231.8+4.2 nebula and its orientation 
towards the observer showing the rotated spheroidal shell geometry of the
lobes explained in the text.}
\label{3dmod}
\end{figure}

\begin{figure}[htb]
\centering
\resizebox{0.9\hsize}{!}{\includegraphics{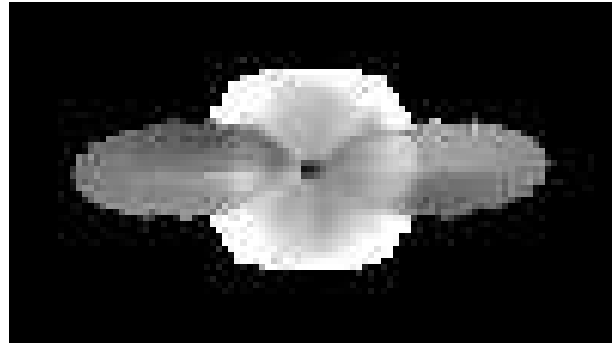}}
\caption{Linear polarization map of the 3-D model nebula consisting of
a pair of triaxial ellipsoidal shells whose major axes almost intersect at the
central illuminating source and an equatorial disc whose thickness along the
bipolar axis is 40\% of the length of each lobe. The structure is tilted
at 40$^\circ$ to the plane of the sky and the bipolar axis is rotated by
40$^\circ$ around the nebula major axis.}
\label{polmod}
\end{figure}

\subsection{New southern knot?}

On the unrotated O and E ray images, there is a stellar (unresolved)
knot on the projected major axis of the nebula at 36$''$ offset from the
position of QX Pup in the direction of the northern lobe. However on 
the 45$^\circ$ rotated images this knot is off the imaged area, so its 
polarization cannot be measured from these data. This could be a star
but its position on the projection of the major axis of the nebula is
suggestive. Although this is the
lower velocity lobe (e.g. Alcolea et al. \cite{alco}), it may be
a high velocity ejectum similar to the knots identified by 
Reipurth (\cite{reip}) in the arc shaped nebula to the south 
of the southern lobe. Follow up emission line imaging and/or
spectrometry are suggested to confirm its association with 
OH~231.8+4.2. 

\section{Conclusions}
A linear polarization map of OH~0231.8+4.2 has been obtained in
the K$_{S}$ band with the SOFI multi-mode instrument. A wealth of structure
is revealed in the polarization and two elongated features in the 
northern bipolar lobe have very high linear polarization. A geometrical
model with bipolar lobes and an equatorial disc is proposed which 
accounts for some of these features. However the high
polarization finger in the northern lobe, and the 
position angle of polarization vectors in the surrounding (disc) 
material, are not explained. Higher spatial resolution polarimetry of
OH~231.8+4.2 is achievable from the ground with AO, and 
polarization maps at different wavelengths would further understanding of 
this fascinating target.

\begin{acknowledgements}
We sincerely thank Jean-Gabriel Cuby who was not only responsible for the
fine functioning of the instrument but also obtained all the observations.
We are very grateful to Alan Moorwood for master minding the SOFI instrument 
and for prompting us to provide a calibration target; the one we
chose also turned out to be scientifically interesting.
\end{acknowledgements}

\end{document}